\documentclass[12pt]{article}

\usepackage{amsmath,amssymb}
\allowdisplaybreaks[4]

\begin{document}

\bibliographystyle{unsrt}

\begin{titlepage}

\title{{\bf Covariant Poisson equation with compact Lie algebras} \\ Existence and
smoothness of solutions}
\author{Antti Salmela\footnote{Email: Antti.Salmela@Helsinki.Fi} \\
\it Theoretical Physics Division \\ \it Department of Physical Sciences \\
\it P.O. Box 64, 00014 University of Helsinki, Finland}
\date{}

\maketitle

\thispagestyle{empty}

\begin{abstract}
The covariant Poisson equation for Lie algebra -valued mappings defined on
$\mathbb{R}^3$ is studied using functional analytic methods. Weighted covariant Sobolev
spaces are defined and used to derive sufficient conditions for the existence and
smoothness of solutions to the covariant Poisson equation. These conditions require,
apart from suitable continuity, appropriate local $L^p$ integrability of the gauge
potentials and global weighted $L^p$ integrability of the curvature form and the source.
The possibility of nontrivial asymptotic behaviour of a solution is also considered. As a
by-product, weighted covariant generalisations of Sobolev embeddings are established.
\end{abstract}

\end{titlepage}

\pagebreak

\newenvironment{ass}{\hfill \vspace{12pt plus 5pt minus 3pt} \break \bf Assumption \it}
{\rm \hfill \vspace{-12pt plus -5pt minus -3pt} \break}
\newtheorem{theorem}{Theorem}[section]{\bf}{\it}
\newtheorem{proposition}{Proposition}[section]{\bf}{\it}
\newtheorem{lemma}{Lemma}[section]{\bf}{\it}
\newenvironment{proof}{\it Proof. \rm}{\hfill \vspace{8pt plus 3pt minus 1.5pt} \break}
\newenvironment{midproof}{\it Proof. \rm}{}
\newenvironment{noproof}{}{\hfill \vspace{8pt plus 3pt minus 1.5pt} \break}
\newenvironment{acknowledgements}{\hfill \vspace{12pt plus 5pt minus 3pt} \break \bf
Acknowledgements \rm \hfill \vspace{8pt plus 3pt minus 1.5pt} \break}{}
\renewcommand{\theequation}{\arabic{section}.\arabic{equation}}

\section{Introduction}

Let us consider two mappings $Z$ and $F$ taking values in some compact Lie algebra
$\cal G$. The covariant generalisation of Poisson's equation on $\mathbb{R}^3$ is then
\begin{equation}\label{kala}
\Delta(A) Z = F,
\end{equation}
where
\begin{eqnarray*}
&& \Delta(A) = \sum_{k=1}^3 \nabla_k^2, \\*
&& \nabla_k = \partial_k + [A_k, \, \cdot \,]
\end{eqnarray*}
and $A_k$ stands for the $\cal G$-valued gauge potential. This equation arises
frequently in gauge theories and it is mostly connected with Gauss's law. In the
Lagrangian formulation of Yang--Mills theories this connection is explicit, since
Gauss's law reads
$$ \Delta(A) A_0 = \sum_{k=1}^3 \nabla_k \dot{A}_k - J_0, \qquad \dot{A}_k =
\frac{\partial A_k}{\partial t},$$
$J_0$ denoting the matter density. In the
Hamiltonian formalism the Gauss law becomes a divergence equation
$$ \sum_{k=1}^3 \nabla_k E_k = J_0. $$
If we split the colour-electric field $E_k$ into longitudinal and transverse components
by
\begin{eqnarray*}
&& E_k = E_k^L + E_k^T, \\*
&& \sum_{k=1}^3 \nabla_k E_k^L = J_0, \qquad \sum_{k=1}^3 \nabla_k
E_k^T = 0,
\end{eqnarray*}
then a solution for the longitudinal component is given by
$$ E_k^L = \nabla_k \Phi, $$
where $\Phi$ satisfies the covariant Poisson equation
$$ \Delta(A) \Phi = J_0. $$
In addition to these cases the Poisson equation also arises if one tries to transform
an arbitrary gauge potential ${\cal A}_{\mu}$ into an equivalent potential
$$ A_{\mu} = \omega^{-1} {\cal A}_{\mu}  \omega + \omega^{-1} \partial_{\mu} \omega, $$
where $A_{\mu}$ satisfies the generalised Coulomb gauge condition
$$ \sum_{k=1}^3 \nabla_k \dot{A_k} = 0, $$
which was proposed by Cronstr\"{o}m in ref. \cite{ccg}. It turns out that the gauge
transformation matrix $\omega$ takes the form of a time-ordered exponential
$$ \omega(x^0, \mathbf{x}) = \left[ T \exp \bigl( - \int_0^{x^0} \hspace{-0.5em} Z(\tau,
\mathbf{x}) \, d\tau \bigr) \right] \omega(0, \mathbf{x}), $$
where $Z$ satisfies the
Poisson equation
$$ \Delta({\cal A}) Z = \sum_{k=1}^3 \nabla_k({\cal A}) \dot{{\cal A}}_k. $$
These are the most important physical reasons for studying this equation. From a purely
mathematical point of view the covariant Poisson equation is also interesting, because
it is a special case of an elliptic system of partial differential equations which, to
the best of my knowledge, has not been considered before in the literature. Research
has been done regarding solutions of the whole set of classical Yang--Mills equations
\cite{s,gv,cbc,em}, but these approaches lead to hyperbolic evolution equations where
Gauss's law becomes a constraint which must hold at all times. The Cauchy data of the
problem is assumed to satisfy the constraint and the evolution equations are used to
show that the constraint then persists. The starting point of this paper is quite
different, because time evolution is not present in the covariant Poisson equation.
Time serves only as a parameter of the mappings in question, but no specific evolution
equations are required to hold.

The line of thought and the most important results of this paper are included in the
three sections to come. The second section deals with the existence of a distributional
solution to the covariant Poisson equation. Conditions for making this solution
smoother are derived in the third section, and the possibility of constructing a
solution formula is touched upon in the concluding fourth section. Proofs are gathered
in the last section, but the preceding sections are meant to be comprehensible even
without reference to the proofs. My notations and conventions are summarised below. Gauge
potentials are written as
$$ A_k(x) = A_k^a(x) T_a, $$
where the Lie algebra generators $T_a$ satisfy
$$ [T_a, T_b] = {f_{ab}}^c T_c. $$
Summation over repeated indices is implied. The components $A_k^a$ and the structure
constants ${f_{ab}}^c$ are real, because all compact Lie algebras are real. The
curvature form of the gauge potential is defined as
$$ G_{kl} = \partial_l A_k - \partial_k A_l - [A_k,A_l], $$
and the covariant divergence of $G_{kl}$ is
$$ (\nabla \cdot G)_k = \sum_{j=1}^3 \nabla_j G_{jk}. $$
The definition of a compact Lie algebra implies that there exists a positive definite
inner product
$$ (X,Y) = h_{ab} X^a Y^b $$
which is invariant under the adjoint action of the gauge group. The infinitesimal form
of this property states that
\begin{equation} \label{impro}
(X,[Y,Z]) = - ([Y,X],Z) \qquad \text{for all} \hspace{0.6em} X, Y, Z \in {\cal G}.
\end{equation}
If $\cal G$ is semisimple, this inner product can be chosen to be the negative of the
Killing form (i.e. $h_{ab} = - {f_{ac}}^d {f_{bd}}^c$). The norm induced by this inner
product will be denoted by $|\cdot|$. More generally, for ${\cal G}$-valued tensors
$X_{i_1 \cdots i_r}$ the norm with $n$th covariant derivatives reads
$$ |\nabla^n X| = \left( \sum_{k_1} \cdots \sum_{k_n} \sum_{i_1} \cdots \sum_{i_r}
| \nabla_{k_1} \cdots \nabla_{k_n} X_{i_1 \cdots i_r} |^2 \right)^{1/2}, $$
and the norm $|\partial^n X|$ with ordinary derivatives is defined analogously. Within
the context of $L^p$ norms it is always understood that the notation $||X||_p$ stands for
$|| \, |X| \, ||_p$. We will also need multi-index notations for the covariant
derivatives. Due to the noncommuting nature of these derivatives it is appropriate to
deviate here from the usual definition and instead let a multi-index $\alpha$ denote an
$n$-tuple of ordered coordinate indices
$$ \alpha = (k_1, \dots, k_n) $$
with the notation for covariant derivatives
$$ \nabla^{\alpha} X = \nabla_{k_1} \cdots \nabla_{k_n} X. $$
The partial derivatives $\partial^{\alpha}$ are defined similarly. The order of a
multi-index is simply $|\alpha| = n$, and the symbol $\nabla^n$ stands for the
collection of all covariant derivatives $\nabla^{\alpha}$ of order $n$. At this point
the reader should take care so as not to confuse the collection of second order
covariant derivatives $\nabla^2$ with the covariant Laplacian $\Delta(A)$, and the same
warning also applies to the norms $|\nabla^2 X|$ and $|\Delta(A) X|$ which have
different meanings. Throughout the paper I will denote positive constants by $C$, but
the values of the constants may change from line to line as they are inessential for the
proofs.

\section{Existence of a weak solution}

\setcounter{equation}{0}

A weak solution of the covariant Poisson equation (\ref{kala}) satisfies
$$ - \int \sum_{k=1}^3 (\nabla_k \widetilde{\Phi}, \nabla_k Z) \, d^3 x = \int
(\widetilde{\Phi}, F) \, d^3 x \qquad \text{for all} \hspace{0.6em} \widetilde{\Phi}
\in C^{\infty}_c(\mathbb{R}^3, {\cal G}),$$
when the weak covariant derivatives
$\nabla_k Z$ and their commutators with $A_k$ are locally integrable. The subscript $c$
above refers to compact support. In order to control the asymptotic behaviour of $Z$ in
subsequent calculations I introduce weights by writing
$$ \widetilde{\Phi}(x) = \frac{1}{w(x)^{1-\sigma}} \Phi(x), \quad w(x) = \left( 1 +
|x|^2 \right)^{1/2}, \quad 0 < \sigma \leq 1. $$
This leads to an equivalent definition
\begin{eqnarray}
&& \int \sum_{k=1}^3 \frac{1}{w^{1-\sigma}} (\nabla_k \Phi, \nabla_k Z) \, d^3 x  \, -
\, (1-\sigma) \int \sum_{k=1}^3 \frac{x_k}{w^{3-\sigma}} (\Phi, \nabla_k Z) \, d^3 x
\nonumber \\*
&& \hspace{2em} = - \int \frac{1}{w^{1-\sigma}} (\Phi, F) \, d^3 x \qquad \text{for
all} \hspace{0.6em} \Phi \in C^{\infty}_c(\mathbb{R}^3, {\cal G}). \label{weak}
\end{eqnarray}
The proof that there exists a solution to this equation is based on the following
well-known theorem:
\begin{theorem}[Lax--Milgram]
Let $H$ be a real Hilbert space with norm $||\cdot||$ and let $B$ denote a bilinear
mapping $B: H \times H \rightarrow \mathbb{R}$. Assume that there exist constants
$\alpha$, $\beta > 0$ such that
\begin{subequations}
\begin{eqnarray}
&& |B(Y,Z)| \leq \alpha \, ||Y|| \, ||Z|| \qquad \text{for all} \hspace{0.6em} Y, Z \in H
\label{lm1} \\*
&& B(Z,Z) \geq \beta \, ||Z||^2. \label{lm2}
\end{eqnarray}
\end{subequations}
If $f$ is a bounded linear functional on $H$, then there exists in $H$ a unique element
$Z$ such that
$$ B(Y,Z) = f[Y] \qquad \text{for all} \hspace{0.6em} Y \in H. $$
\end{theorem}
\begin{noproof}
For a proof, see e.g. ref. \cite{e}, section 6.2.1.
\end{noproof}
In our case the mapping $B(\Phi,Z)$ is given by the left hand side of the
definition (\ref{weak}) and the functional $f$ by the right hand side. A suitably
defined Sobolev space will play the role of the Hilbert space $H$. We start by defining
an inner product on the space $C^1_c(\mathbb{R}^3, {\cal G})$ by
\begin{equation}\label{tulo}
<\Phi,\Psi>_1 = \int \frac{1}{w^{1-\sigma}} \sum_{k=1}^3 (\nabla_k \Phi, \nabla_k \Psi)
\, d^3 x,
\end{equation}
where it is assumed that $A_k \in L^2_{loc}(\mathbb{R}^3, {\cal G})$. The norm
associated with this inner product will be denoted by $||\cdot||_{1,2}$. In order to
check that this formula really defines an inner product we need the following two
results:
\begin{lemma}\label{kommu}
Let ${\cal G}$ be a compact Lie algebra. Then there exists a constant $C>0$ such that
$$ |\, [X,Y] \, | \leq C |X| \, |Y| $$
for all elements $X, Y \in {\cal G}$.
\end{lemma}
\begin{midproof}
See section \ref{kommupr}.
\end{midproof}
\begin{proposition}\label{goin}
Let $A_k \in L^2_{loc}(\mathbb{R}^3, {\cal G})$. Then the inequality
\begin{equation} \label{goineq}
\left( \int \frac{1}{w^{3-\sigma}} |\Phi|^2 \, d^3 x \right)^{1/2} \leq C \left( \int
\frac{1}{w^{1-\sigma}} \sum_{k=1}^3 |\nabla_k \Phi|^2 \, d^3 x \right)^{1/2}
\end{equation}
holds for all mappings $\Phi \in C^1_c(\mathbb{R}^3, {\cal G})$. The constant $C$ does
not depend on $\Phi$.
\end{proposition}
\begin{proof}
See section \ref{goinpr}.
\end{proof}
Lemma \ref{kommu} together with H\"{o}lder's inequality ensures that the expression
(\ref{tulo}) is defined, and Proposition \ref{goin} guarantees that the norm
$||\Phi||_{1,2}$ vanishes only when $\Phi \equiv 0$. The first order Sobolev space is
now defined as the completion of $C^1_c(\mathbb{R}^3, {\cal G})$ in the norm
$||\cdot||_{1,2}$ i.e.
$$ H_1(\mathbb{R}^3, {\cal G}) = \overline{C^1_c(\mathbb{R}^3, {\cal G})}. $$
It goes without saying that this Sobolev space depends on $\sigma$, which is kept fixed
in forthcoming calculations. Considering Cauchy sequences in \break $H_1(\mathbb{R}^3,
{\cal G})$ with elements of class $C^1_c(\mathbb{R}^3, {\cal G})$ we can extend the inner
product (\ref{tulo}) and Proposition \ref{goin} to $H_1(\mathbb{R}^3, {\cal G})$. As a
result, we see that the norm $||\cdot||_{1,2}$ can be extended to $H_1(\mathbb{R}^3,
{\cal G})$ as it only vanishes for mappings equivalent to zero. Let us now apply the
Lax--Milgram theorem to the bilinear mapping $B$ of equation (\ref{weak}). Employing
Schwarz's inequality and Proposition \ref{goin} yields
$$ |B(Y,Z)| \leq \left[ 1 + C(1-\sigma) \right] ||Y||_{1,2} \, ||Z||_{1,2} \qquad
\text{for all} \hspace{0.6em} Y, Z \in H_1(\mathbb{R}^3, {\cal G}), $$
which
establishes the property (\ref{lm1}). Coercivity is proved by examining the second term
of $B$ with mappings of class $C^1_c(\mathbb{R}^3, {\cal G})$ first. Equation
(\ref{impro}) shows that almost everywhere
$$ (\Phi, \nabla_k \Phi) = (\Phi, \partial_k \Phi) = \frac{1}{2} \partial_k |\Phi|^2, $$
and this leads to the identity
\begin{eqnarray*}
\lefteqn{- (1-\sigma) \int \sum_{k=1}^3 \frac{x_k}{w^{3-\sigma}} (\Phi, \nabla_k \Phi) \, d^3 x
=} \\*
&& - \frac{1}{2} (1-\sigma) \int \sum_{k=1}^3 \partial_k \left( \frac{x_k}{w^{3-\sigma}}
|\Phi|^2 \right) \, d^3 x + \frac{1}{2} (1-\sigma) \int \frac{3 + \sigma |x|^2}{w^{5-
\sigma}} |\Phi|^2 \, d^3 x.
\end{eqnarray*}
The first integral on the right can be converted into a surface integral which vanishes
due to the compact support of $\Phi$. The second integral is non-negative and hence
$$ - (1-\sigma) \int \sum_{k=1}^3 \frac{x_k}{w^{3-\sigma}} (\Phi, \nabla_k \Phi) \, d^3
x \geq 0. $$ Taking a Cauchy sequence $(\Phi_m)$ of mappings belonging to
$C^1_c(\mathbb{R}^3, {\cal G})$ and converging to an arbitrary element $Z$ in
$H_1(\mathbb{R}^3, {\cal G})$ we see that the inequality above holds for $Z$ also.
Indeed, the integrals with $\Phi_m$ converge to the integral with $Z$ by virtue of
Proposition \ref{goin} and Schwarz's inequality. As a result,
$$ B(Z,Z) \geq ||Z||_{1,2}^2 \qquad \text{for all} \hspace{0.6em} Z \in H_1(\mathbb{R}^3,
{\cal G})$$
and the condition (\ref{lm2}) is satisfied. Now the Lax--Milgram theorem
yields a solution to equation (\ref{weak}) whenever the right hand side defines a
bounded linear functional on $H_1(\mathbb{R}^3, {\cal G})$. By Schwarz's inequality the
condition
\begin{equation}\label{fcon}
|| w^{\frac{1}{2}(1+\sigma)} F ||_2 < \infty
\end{equation}
suffices for that purpose. We can summarise the result as follows:
\begin{theorem}\label{exist}
Suppose that $A_k \in L^2_{loc}(\mathbb{R}^3, {\cal G})$ and the condition (\ref{fcon})
is satisfied. Then there exists in $H_1(\mathbb{R}^3, {\cal G})$ a weak solution of the
covariant Poisson equation (\ref{kala}).
\end{theorem}
Some words about the asymptotic behaviour of the solution are appropriate at this
stage. Roughly speaking, the mappings of $H_1(\mathbb{R}^3, {\cal G})$ tend to zero at
infinity like some negative power of the radius $|x|$. In fact, there is a more general
theorem by Cronstr\"{o}m \cite{cc} stating that two classical solutions of the
covariant Poisson equation are equal if their difference vanishes at infinity. Thus in
order to obtain more solutions we must consider mappings with nontrivial behaviour at
infinity. Let us assume that a mapping $Z_0$ solves the Poisson equation asymptotically
in the sense that
\begin{equation}\label{nontriv}
|| w^{\frac{1}{2}(1+\sigma)} \left( F - \Delta(A) Z_0 \right)||_2 < \infty.
\end{equation}
Then by Theorem \ref{exist} there exists a solution $Y \in H_1(\mathbb{R}^3, {\cal G})$
to the equation
$$ \Delta(A) Y = F - \Delta(A) Z_0 $$
and accordingly, a mapping defined by
\begin{equation}\label{yzo}
Z = Y + Z_0
\end{equation}
solves the original equation (\ref{kala}). Since the mapping $Y$ tends to zero at
infinity in a generalised sense, the asymptotic behaviour of $Z$ is now determined by
$Z_0$. Solutions with nontrivial behaviour can thus be obtained by constructing such a
mapping $Z_0$ that the condition (\ref{nontriv}) is in force. To a great extent the
details of the construction depend on the asymptotic behaviour of the source $F$.

\section{Smoothness of solutions}

\setcounter{equation}{0}

Now that we have proved the existence of a distributional solution to the Poisson
equation, it is time to consider the smoothness properties of this solution. The
standard technique is to define higher order Sobolev spaces and then apply the fact
that these spaces are continuously embedded in the spaces of mappings with continuous
and bounded derivatives. Then it remains to derive a suitable \textit{a priori}
estimate which enables us to conclude that a solution under certain assumptions belongs
to a higher order Sobolev space. Let us begin with the definitions of $n$th order Sobolev
spaces. Throughout this section the following local assumptions are supposed to hold:
\begin{ass}
\begin{subequations}\label{asmo}
If $n=2$, we assume that
\begin{eqnarray} \label{asmo1}
&& A_k \in C(\mathbb{R}^3, {\cal G}), \quad \partial^1 A_k \in L^3_{loc}(\mathbb{R}^3,
{\cal G}), \quad \partial^2 A_k \in L^1_{loc}(\mathbb{R}^3, {\cal G}), \nonumber \\*
&& G_{kl} \in C(\mathbb{R}^3, {\cal G}),
\end{eqnarray}
while in the case $n \geq 3$ these assumptions are replaced by
\begin{eqnarray} \label{asmo2}
&& A_k \in C^{n-3}(\mathbb{R}^3, {\cal G}), \quad \partial^{n-2} A_k \in
L^q_{loc}(\mathbb{R}^3, {\cal G}) \hspace{0.6em} \text{for some} \hspace{0.6em} q >3,
\nonumber \\*
&& \partial^{n-1} A_k \in L^2_{loc}(\mathbb{R}^3, {\cal G}), \quad G_{kl} \in
C^{n-3}(\mathbb{R}^3, {\cal G}).
\end{eqnarray}
\end{subequations}
\end{ass}
These conditions enable us to define a higher order inner product on the space
$C^n_c(\mathbb{R}^3, {\cal G})$ by
$$ <\Phi,\Psi>_n = \sum_{p=1}^n \int w^{(2p-3)(1-\sigma)} \sum_{k_1=1}^3 \cdots
\sum_{k_p=1}^3 (\nabla_{k_1} \cdots \nabla_{k_p} \Phi, \nabla_{k_1} \cdots \nabla_{k_p}
\Psi) \, d^3 x. $$
The corresponding norm will be denoted by $||\cdot||_{n,2}$. Now the
$n$th order Sobolev space is defined as
$$ H_n(\mathbb{R}^3, {\cal G}) = \overline{C^n_c(\mathbb{R}^3, {\cal G})}, $$
where the closure is taken in the norm $||\cdot||_{n,2}$. These spaces are going to be
embedded in the spaces $C_B^{n-2}(\mathbb{R}^3, {\cal G})$ consisting of
$C^{n-2}(\mathbb{R}^3, {\cal G})$ mappings for which the norm
$$ ||\Phi||_{n-2,\infty} = \max_{0 \, \leq p \, \leq n-2} \, \sup_{x \in \mathbb{R}^3}
\left\{ w(x)^{p(1-\sigma)} |\nabla^p \Phi(x)| \right\} $$
is finite. Note that $A_k \in
C^{n-3}(\mathbb{R}^3, {\cal G})$ at this stage and the covariant derivatives in this
definition thus yield continuous mappings. We can now formulate the embedding property
exactly:
\begin{theorem}[Sobolev embeddings]
Suppose that the assumptions \hfil \break (\ref{asmo}) hold. Then the spaces
$H_n(\mathbb{R}^3, {\cal G})$ are continuously embedded in the spaces
$C_B^{n-2}(\mathbb{R}^3, {\cal G})$ i.e.
\begin{equation} \label{embed}
H_n(\mathbb{R}^3, {\cal G}) \subset C_B^{n-2}(\mathbb{R}^3, {\cal G}).
\end{equation}
\end{theorem}
\begin{proof}
See section \ref{bedpr}.
\end{proof}
The following \textit{a priori} estimate establishes the inductive chain which allows
us to include the distributional solution into higher order Sobolev spaces:
\begin{proposition}\label{prior}
Suppose that the conditions (\ref{asmo}) are in force. Then the inequality
\begin{equation} \label{apri}
||\Phi||_{n,2} \leq C \left( ||\Phi||_{n-1,2} + ||w^{(n-3/2)(1-\sigma)} \nabla^{n-2}
\Delta(A) \Phi||_2 \right)
\end{equation}
holds for all $\Phi \in C^{n+1}_c(\mathbb{R}^3, {\cal G})$. The constant $C$ is
independent of $\Phi$, but it depends on weighted $L^p$ norms of the curvature form as
follows:
$$ C = C \left( ||w^{2(1-\sigma)} G||_{\infty}, ||w^{2(1-\sigma)} (\nabla \cdot G)||_3
\right), $$
when $n=2$, whereas
$$ C = C \left( ||w^{(p+2)(1-\sigma)} \nabla^p
G||_{\infty}, ||w^{(p+2)(1-\sigma)} \nabla^p (\nabla \cdot G)||_3 \right), \quad p = 0,
\dots, n-3 $$
when $n \geq 3$. The constant is finite when the norms in question are
finite.
\end{proposition}
\begin{proof}
See section \ref{priorpr}.
\end{proof}
It should be mentioned that the form of this estimate is not unique as there is some
freedom in choosing the curvature norms which are to be bounded. My guideline has been
to keep the asymptotic requirements on the curvature as mild as possible and to make
the transition to the unweighted case $\sigma = 1$ easy. It is possible to relax the
local assumptions, but that might happen at the cost of asymptotic conditions. The
reason why I am worrying about the asymptotic behaviour of the curvature form is a
paper by Coleman \cite{sc} in which he shows that the only non-singular solution of the
sourcefree Yang--Mills equations is the vacuum solution provided that the condition
$$ \lim_{|x| \rightarrow \infty} |x|^{3/2 + \epsilon} G_{\mu \nu}^a(x,t) = 0, \qquad 0 <
\epsilon < \frac{1}{2} $$
holds uniformly in time in the region $t > 0$. It is easy to
check that the norms of Proposition \ref{prior} escape this condition since $\sigma=1$
in the sourcefree case. When sources are present, the solution will always be
nontrivial. Now in order to extend Proposition \ref{prior} to mappings in
$H_n(\mathbb{R}^3, {\cal G})$ we employ the following density result:
\begin{lemma}\label{jono}
Suppose that the conditions (\ref{asmo}) are in force. If
\begin{equation}\label{oljo}
Z \in H_{n-1}(\mathbb{R}^3, {\cal G}) \quad \textrm{and} \quad w^{(n-3/2)(1-\sigma)}
\nabla^{n-2} \Delta(A) Z \in L^2(\mathbb{R}^3, {\cal G}),
\end{equation}
then there exists a sequence $(\Phi_m)$ of mappings belonging to
$C^{\infty}_c(\mathbb{R}^3, {\cal G})$ such that $\Phi_m \rightarrow Z$ in
$H_{n-1}(\mathbb{R}^3, {\cal G})$ and also
$$ ||w^{(n-3/2)(1-\sigma)} \nabla^{n-2} \Delta(A) (Z - \Phi_m)||_2 \longrightarrow 0. $$
\end{lemma}
\begin{proof}
See section \ref{jonopr}.
\end{proof}
Taking a Cauchy sequence $(\Phi_m)$ of $C^{\infty}_c(\mathbb{R}^3, {\cal G})$ mappings
converging to $Z$ in both norms of Lemma \ref{jono} and inserting it into the estimate
(\ref{apri}) shows that the sequence also converges in $H_n(\mathbb{R}^3, {\cal G})$
towards $Z$. As a result,
$$ Z \in H_n(\mathbb{R}^3, {\cal G}) \subset C_B^{n-2}(\mathbb{R}^3, {\cal G}). $$
If the derivatives of $Z$ satisfy the Poisson equation (\ref{kala}) in a distributional
sense, the covariant Laplacian norms of $Z$ can then be replaced by the corresponding
norms of $F$. In short:
\begin{theorem}\label{smooth}
Suppose that the assumptions (\ref{asmo}) hold and the curvature norms of Proposition
\ref{prior} are finite. If in addition
\begin{equation}\label{fadcon}
||w^{(p+1/2)(1-\sigma)} \nabla^p F||_2 < \infty, \quad p = 0, \dots, n-2,
\end{equation}
then the solution of the covariant Poisson equation given by Theorem \ref{exist}
belongs to $C_B^{n-2}(\mathbb{R}^3, {\cal G})$.
\end{theorem}
When solutions with nontrivial asymptotic behaviour are constructed by writing $Z$ in
the form (\ref{yzo}), this theorem can be used to deduce the smoothess properties of
$Y$, replacing only $F$ with $F - \Delta(A) Z_0$ in the condition (\ref{fadcon}). The
ultimate smoothness of $Z$ then depends on the properties of $Z_0$.

\section{Conclusions}

\setcounter{equation}{0}

Theorems \ref{exist} and \ref{smooth} are the main results regarding solutions of the
covariant Poisson equation. All the assumptions of these theorems are only sufficient,
and it remains an open question to determine the necessary conditions. However, it is
likely that some bounds should be imposed on the curvature form. In fact, similar
studies by Aubin \cite{a}, Cantor \cite{ca} and Eichhorn \cite{ei} on Riemannian
manifolds and vector bundles indicate that bounds are also needed for the curvature
tensor and its derivatives in these approaches. As regards the smoothness assumptions
(\ref{asmo}), the fact that they are all local is consistent with the principles of
fibre bundle theory in the sense that gauge potentials are considerer as local objects.
Global integrability conditions can be imposed on the curvature form without problems
because the Lie algebra norms $|\nabla^p G|$ and $|\nabla^p(\nabla \cdot G)|$ are gauge
invariant.

A slight shortcoming of the Lax--Milgram theorem is the fact that it does not give an
explicit formula for the solution whose existence it proves. A formal procedure exists
for separable Hilbert spaces, though. Following the proof of separability for ordinary
Sobolev spaces (see e.g. ref. \cite{mac}, section 6.3) it is straightforward to check
that the weighted covariant Sobolev space $H_1(\mathbb{R}^3, {\cal G})$ is indeed
separable. Then it is possible, at least in principle, to construct a countable
orthonormal basis $\{ \Psi_n \}$ for $H_1(\mathbb{R}^3, {\cal G})$. Expressing $Z$ and
$\Phi$ in equation (\ref{weak}) as generalised Fourier series
$$ Z = \sum_{n=1}^{\infty} a_n \Psi_n, \qquad \Phi = \sum_{n=1}^{\infty} b_n \Psi_n $$
we see that the coefficients $a_n$ are to be determined from the equation
$$ a_n - \sum_{m=1}^{\infty} a_m (1-\sigma) \int \sum_{k=1}^3 \frac{x_k}{w^{3-\sigma}}
(\Psi_n, \nabla_k \Psi_m) \, d^3 x = - \int \frac{1}{w^{1-\sigma}} (F, \Psi_n) \, d^3
x. $$ The superficial simplicity of this solution formula hides the practical
difficulties in constructing an orthonormal basis for $H_1(\mathbb{R}^3, {\cal G})$.
Also in traditional approaches to the covariant Poisson equation the solution formulas
tend to be too complicated. Employing suitable asymptotic conditions and inverting the
ordinary Laplacian it is possible to convert the equation into an integral equation. No
matter whether one iterates the resulting equation or applies Fredholm's formulas, the
solution will always be rather complicated. Maybe we should not even expect the
covariant Poisson equation to have simple and elegant solutions.

\section{Proofs}

\setcounter{equation}{0}

\subsection{Lemma \ref{kommu}}\label{kommupr}

This result is almost self-evident, because the structure constants are \break bounded.
Anyway, let us be explicit and consider the norm of the commutator
\begin{eqnarray*}
|\, [X,Y] \, |^2 &=& X^a M_{ae} X^e, \\*
M_{ae} &=& h_{cd} {f_{ab}}^c {f_{ef}}^d Y^b Y^f.
\end{eqnarray*}
Using Cauchy's inequality
$$ X^a X^e \leq \frac{1}{2} \left[ \left( X^a \right)^2 + \left( X^e \right)^2 \right]
$$
we can derive the bound
$$ |\, [X,Y] \, |^2 \leq C(Y) \sum_{a=1}^d \left( X^a \right)^2, $$
where
$$ C(Y) = \max_{1 \leq a \leq d} \left\{ \sum_{e = 1}^d |M_{ae}| \right\} $$
and $d$ denotes the dimension of the algebra. Applying
similar techniques to the matrix $M_{ae}$ we can also bound its elements by
$$ |M_{ae}| \leq \widetilde{C}_{ae} \sum_{b=1}^d \left( Y^b \right)^2, $$
where the matrix $\widetilde{C}_{ae}$ is independent of $Y$. Hence we obtain an
identical bound for the constant $C(Y)$ above. The proof is completed by noting that
the metric tensor $h_{ab}$ has strictly positive eigenvalues. If $\lambda_{min}$ and
$\lambda_{max}$ denote the smallest and largest eigenvalues of $h$, we have
$$ \lambda_{min} \sum_{a=1}^d \left( X^a \right)^2 \leq |X|^2 \leq \lambda_{max}
\sum_{a=1}^d \left( X^a \right)^2 $$
i.e. the norm $|\cdot|$ is equivalent with the
$d$-dimensional Euclidean norm. $\Box$

\subsection{Proposition \ref{goin}}\label{goinpr}

Let us begin with a more general theorem due to Gurka and Opic \cite{go} regarding real
functions on $\mathbb{R}^3$.
\begin{theorem}[Gurka--Opic]
Let $1 \leq p < \infty$ and let $\bar{v}_0$, $\bar{v}_1$ be real, measurable, almost
everywhere positive and finite functions of one real variable. Suppose futher that they
are bounded below and above by positive constants on every compact interval of the
positive real axis and that there exists a constant $k > 0$ and a number $t_0 \in
]0,\infty[$ such that
$$ \bar{v}_0(t) \geq k \, \bar{v}_1(t) t^{-p} \quad \textit{for all} \quad t \geq t_0. $$
Finally, if
\begin{equation}\label{cond}
\sup_{0 < \tau < \infty} \left( \int_0^{\tau} \bar{v}_0(t) t^2 \, dt \right)^{1/p}
\left( \int_{\tau}^{\infty} \left[\bar{v}_1(t) t^2\right]^{-1/(p-1)} dt \right)^{1 -
1/p} < \infty,
\end{equation}
then there exists a constant $C>0$ such that
\begin{equation}\label{apu51}
\left( \int |u(x)|^p \, \bar{v}_0(|x|) \, d^3 x \right)^{1/p} \leq C \left(
\sum_{i=1}^3 \int |\partial_i u(x)|^p \, \bar{v}_1(|x|) \, d^3 x \right)^{1/p}
\end{equation}
for all integrable functions $u$ such that the norms of the inequality are finite. The
constant $C$ is independent of $u$.
\end{theorem}
\begin{proof}
This is a special case of Theorem 14.5. in ref. \cite{go}. In their re\-marks Gurka and
Opic state without proof that conditions (\ref{cond}) and (\ref{apu51}) are actually
equivalent. $\Box$
\end{proof}
We can now make use of this theorem and choose
$$ p = 2, \qquad \bar{v}_0(t) = (1+t^2)^{-\frac{3-\sigma}{2}}, \qquad \bar{v}_1(t) =
(1+t^2)^{-\frac{1-\sigma}{2}}, \qquad 0 < \sigma \leq 1. $$ This leads to the
inequality
\begin{equation} \label{apu52}
\left( \int \frac{1}{w^{3-\sigma}} |u|^2 \, d^3 x \right)^{1/2} \leq C \left( \int
\frac{1}{w^{1-\sigma}} \sum_{k=1}^3 |\partial_k u|^2 \, d^3 x \right)^{1/2},
\end{equation}
which holds in particular for all $u \in C^1_c(\mathbb{R}^3)$. Condition (\ref{cond})
fails for $\sigma = 0$ indicating that inequality (\ref{apu52}) cannot be
improved by setting $\sigma$ to zero. In order to derive inequality (\ref{goineq}) we
make use of a trick employed by Ginibre and Velo in ref. \cite{gv}. Namely, for a
mapping $\Phi \in C^1_c(\mathbb{R}^3, {\cal G})$ we define a family of functions
\begin{equation}\label{trick}
u_{\delta}(x) = \left( |\Phi(x)|^2 + \delta^2 \right)^{1/2} - \delta, \qquad \delta
> 0.
\end{equation}
Clearly $u_{\delta} \in C^1_c(\mathbb{R}^3)$ and
\begin{equation}\label{apu53}
\partial_k u_{\delta} = \left( |\Phi|^2 + \delta^2 \right)^{-1/2} (\partial_k
\Phi,\Phi).
\end{equation}
As before, we note that
\begin{equation}\label{apu54}
(\partial_k \Phi,\Phi) = (\nabla_k \Phi,\Phi)
\end{equation}
whenever $A_k$ is finite. Since $A_k \in L^2_{loc}(\mathbb{R}^3, {\cal G})$, it follows
that $\nabla_k \Phi \in L^2(\mathbb{R}^3, {\cal G})$ and equation (\ref{apu54}) holds
almost everywhere. Combining equations (\ref{apu53}) and (\ref{apu54}) we see that
\begin{equation}\label{apu55}
|\partial_k u_{\delta}| \leq |\nabla_k \Phi|
\end{equation}
almost everywhere. Inequalities (\ref{apu52}) and (\ref{apu55}) finally yield
inequality (\ref{goineq}) in the limit $\delta \rightarrow 0$. $\Box$

\subsection{Sobolev embeddings (\ref{embed})}\label{bedpr}

These embeddings constitute an obvious generalisation of well-known results for real
functions. Sobolev embeddings for Riemannian vector bundles have previously been derived
by Cantor \cite{ca} and Eichhorn \cite{ei}, but the results are not directly applicable
here, because they deal with unweighted inequalities only. For this reason I prefer to
present a derivation of the embeddings (\ref{embed}) here. The most essential part of the
proof consists of deriving weighted generalisations of the ordinary embeddings, and the
rest follows by employing the trick (\ref{trick}). However general the weighted
inequalities of references \cite{go} and \cite{bh} are, they are of little use here,
because inequalities involving $L^{\infty}$ norms are not included in them. For that
reason I am going to proceed using a method presented in chapter VI.6 of ref.
\cite{cdd2}. Yet there is a small improvement in my calculations below compared with
those of ref. \cite{cdd2}. Namely, I have been able to lower the Sobolev space weight
factors of $n$th partial derivatives from $w^{n-\frac{3-\sigma}{2}}$ to
$w^{(n-3/2)(1-\sigma)}$, which in turn yields milder asymptotic conditions in Theorem
\ref{smooth}. Let us begin with the familiar interpolation inequalities for real
functions \cite{n}:
\begin{theorem}[Gagliardo--Nirenberg--Sobolev]\label{inter}
Let $1 \leq q \leq \infty$, $1 \leq r \leq \infty$, $1 \leq p \leq \infty$, $0 \leq a
\leq 1$ and assume that the following relation holds true:
$$ \frac{1}{p} = \frac{1-a}{r} + a \left( \frac{1}{q} - \frac{1}{3} \right). $$
In the case $p = \infty$ assume further that $r < \infty$ and $a < 1$. Now if $r <
\infty$ or $q \geq 3$, then there exists a constant $C$, depending only on $p$, $q$,
and $r$, such that the inequality
\begin{equation}\label{apu61}
||v||_p \leq C \, ||v||_r^{1-a} \, ||\partial v||_q^a
\end{equation}
holds for all functions $v \in L^r$ whose derivatives lie in $L^q$. In the case $r =
\infty$, $q < 3$ the inequality holds for functions which in addition tend to zero at
infinity or which lie in $L^{r_0}$ for some finite $r_0 > 0$.
\end{theorem}
The next step is to perform a change of variables by
$$ y_k = \frac{x_k}{w(x)^{1 - \sigma}}, \qquad 0 < \sigma \leq 1. $$
This mapping is one-to-one and smooth in the range of $\sigma$ and we can define a new
function $u$ by
$$ u(y) = v(x(y)). $$
For this change of variables
\begin{eqnarray*}
&&\frac{\partial y_k}{\partial x_l} = M_{kl} \, w(x)^{-(1 - \sigma)}, \qquad  M_{kl} =
\delta_{kl} - (1 - \sigma) \frac{x_k x_l}{1 + |x|^2}, \\
&&\det M = \frac{1 + \sigma |x|^2}{1 + |x|^2}, \qquad \sigma \leq \det M \leq 1.
\end{eqnarray*}
Making use of these properties we see that
$$ \int |u(y)|^r \, d^3 y = \int |v(x)|^r (\det M) \, w(x)^{-3(1 - \sigma)} \, d^3 x. $$
As a result, we obtain the inequalities
\begin{equation}\label{apu62}
\sigma^{1/r} ||w^{-\frac{3}{r}(1 - \sigma)} v||_r \leq ||u||_r \leq ||w^{-\frac{3}{r}(1
- \sigma)} v||_r.
\end{equation}
This is also valid if $r = \infty$, because the norms of $u$ and $v$ are equal then.
For the derivative norm we need to know in addition that
\begin{eqnarray*}
&&|\partial u|^2 = \sum_{k,\, l=1}^3 \frac{\partial v}{\partial x_k} \left[ M^{-1}
\left( M^{-1} \right)^T \right]_{kl} \frac{\partial v}{\partial x_l} \, w(x)^{2(1 -
\sigma)}, \\
&&\left[ M^{-1} \left( M^{-1} \right)^T \right]_{kl} = \delta_{kl} + \left[ \left(
\frac{1 + |x|^2}{1 + \sigma |x|^2} \right)^2 - 1 \right] \frac{x_k x_l}{|x|^2}, \\
&& |\xi|^2 \leq \sum_{k,\, l=1}^3 \xi_k \left[ M^{-1} \left( M^{-1} \right)^T
\right]_{kl} \xi_{\,l} \leq \frac{1}{\sigma^2} |\xi|^2 \qquad \textrm{for all} \quad
\xi \in \mathbb{R}^3.
\end{eqnarray*}
These properties together with the previous ones enable us to establish the inequality
\begin{equation}\label{apu63}
\sigma^{1/q} ||w^{(1 - \frac{3}{q})(1 - \sigma)} \partial v||_q \leq ||\partial u||_q
\leq \frac{1}{\sigma} ||w^{(1 - \frac{3}{q})(1 - \sigma)} \partial v||_q.
\end{equation}
Combining the inequalities (\ref{apu61}) -- (\ref{apu63}) we are finally led to
$$ ||w^{-\frac{3}{p}(1 - \sigma)} v||_p \leq C \sigma^{-a-1/p}  \, ||w^{-\frac{3}{r}(1 -
\sigma)} v||_r^{1-a} \, ||w^{(1 - \frac{3}{q})(1 - \sigma)} \partial v||_q^a. $$
We can
still go a step further by defining a new function $u$ through
$$ v(x) = w(x)^{\beta} u(x), \qquad \beta \in  \mathbb{R}. $$
It is easily seen that
$$ |\partial v|^2 \leq \left( w^{\beta} |\partial u| + |\beta| \, w^{\beta - 1} |u|
\right)^2, $$ and applying Minkowski's inequality we get
\begin{eqnarray*}
||w^{\beta - \frac{3}{p}(1 - \sigma)} u||_p &\leq& C \, ||w^{\beta - \frac{3}{r}(1 -
\sigma)}
u||_r^{1-a} \Bigl( ||w^{\beta + (1 - \frac{3}{q})(1 - \sigma)} \partial u||_q \\*
&& \hspace{9.0em} + |\beta| \, ||w^{\beta -1 + (1 - \frac{3}{q})(1 - \sigma)} u||_q
\Bigr)^a.
\end{eqnarray*}
Now the trick (\ref{trick}) allows us to pass over to Lie algebra valued mappings and
to replace $|u|$ by $|\Phi|$ and $|\partial u|$ by $|\nabla \Phi|$ in the inequality
above. In addition to that, it is only required that the weak derivatives of
$u_{\delta}$ are given by the formula
$$\partial_k u_{\delta} = \left( |\Phi|^2 + \delta^2 \right)^{-1/2} (\nabla_k \Phi,
\Phi).$$
Following the proof by Ginibre and Velo \cite{gv} it suffices to assume that
there exists a sequence $(\Phi)_m$ of continuously differentiable mappings
satisfying
\begin{eqnarray*}
&& \Phi_m \longrightarrow \Phi \quad \text{in} \hspace{0.6em} L^1_{loc}(\mathbb{R}^3,
{\cal G}) \hspace{0.6em} \text{and almost everywhere}, \\
&& \nabla_k \Phi_m \longrightarrow \nabla_k \Phi \quad \text{in} \hspace{0.6em}
L^1_{loc}(\mathbb{R}^3, {\cal G}) \hspace{0.6em}
\end{eqnarray*}
in order to guarantee the validity of the derivative formula. In particular, $\Phi$
does not have to be continuous --- local integrability of $\Phi$ and its covariant
derivatives suffices. Using standard smoothing techniques (see section \ref{jonopr}) it
is possible to prove that a sequence $(\Phi)_m$ with the desired properties can be
constructed if $A_k$ is continuous. Thus
\begin{eqnarray*}
||w^{\beta - \frac{3}{p}(1 - \sigma)} \Phi||_p &\leq& C \, ||w^{\beta - \frac{3}{r}(1 -
\sigma)} \Phi||_r^{1-a} \Bigl( ||w^{\beta + (1 - \frac{3}{q})(1 - \sigma)} \nabla
\Phi||_q \\*
&& \hspace{9.2em} + \, |\beta| \, ||w^{\beta -1 + (1 - \frac{3}{q})(1 - \sigma)} \Phi||_q
\Bigr)^a.
\end{eqnarray*}
More generally, we can now apply this inequality to $m$th covariant derivatives of
$\Phi$ with the result
\begin{eqnarray*}
||w^{\beta - \frac{3}{p}(1 - \sigma)} \nabla^m \Phi||_p &\leq& C \, ||w^{\beta -
\frac{3}{r}(1 - \sigma)} \nabla^m \Phi||_r^{1-a} \Bigl( ||w^{\beta + (1 -
\frac{3}{q})(1 - \sigma)} \nabla^{m+1} \Phi||_q \\*
&& + \, |\beta| \, ||w^{\beta -1 +
(1 - \frac{3}{q})(1 - \sigma)} \nabla^m \Phi||_q \Bigr)^a,
\end{eqnarray*}
which holds if the norms on the right hand side are finite and the assumptions of
Theorem \ref{inter} are in force. For the Sobolev embeddings we need the following
three applications of this inequality:
\begin{subequations}
\begin{eqnarray}
&& 1) \quad ||w^{(n-2)(1 - \sigma)} \nabla^{n-2} \Phi||_{\infty} \leq C \, ||w^{(n-5/2)
(1 - \sigma)} \nabla^{n-2} \Phi||_6^{1-a} \label{gns1} \\*
&& \hspace{12.7em} \times \, \Bigl( ||w^{(n - 1 - 3/q)(1 - \sigma)} \nabla^{n-1}
\Phi||_q \nonumber \\*
 && \hspace{14.4em} + (n-2) \, ||w^{(n - 2 - 3/q)(1 - \sigma)} \nabla^{n-2} \Phi||_q
 \Bigr)^{a}, \nonumber \\*
 && \hspace{2.2em} q > 3, \qquad a = \frac{q}{3(q-2)}, \nonumber \\
&& 2) \quad ||w^{(n - 5/2)(1 - \sigma)} \nabla^{n-2} \Phi||_6 \leq C \, \Bigl( ||w^{(n -
5/2)(1 - \sigma)} \nabla^{n-1} \Phi||_2 \label{gns2} \\*
&& \hspace{14.8em} + (n-2) \, ||w^{(n - 7/2)(1 - \sigma)} \nabla^{n-2} \Phi||_2 \Bigr),
\nonumber \\
&& 3) \quad ||w^{(n - 1 - 3/q)(1 - \sigma)} \nabla^{n-1} \Phi||_q \leq C \, ||w^{(n-5/2)
(1 - \sigma)} \nabla^{n-1} \Phi||_2^{1-b} \label{gns3} \\*
&& \hspace{14.0em} \times \, \Bigl( ||w^{(n - 3/2)(1 - \sigma)} \nabla^n
\Phi||_2 \nonumber \\*
 && \hspace{15.7em} + (n-1) \, ||w^{(n - 5/2)(1 - \sigma)} \nabla^{n-1} \Phi||_2
 \Bigr)^{b}, \nonumber \\*
&& \hspace{2.2em} q \leq 6, \qquad b = 3 \left( \frac{1}{2} - \frac{1}{q} \right).
\nonumber
\end{eqnarray}
\end{subequations}
Here it is assumed that $\Phi \in C^n_c(\mathbb{R}^3, {\cal G})$ and the conditions
(\ref{asmo}) are satisfied. Combining these inequalities and recalling the definitions
of the norms $||\cdot||_{n,2}$ and $||\cdot||_{n-2,\infty}$ we get
$$ ||w^{(n-2)(1 - \sigma)} \nabla^{n-2} \Phi||_{\infty} \leq C \, ||\Phi||_{n,2}, $$
and hence
$$ ||\Phi||_{n-2,\infty} \leq C \, ||\Phi||_{n,2} \qquad \text{for
all} \hspace{0.6em} \Phi \in C^n_c(\mathbb{R}^3, {\cal G}). $$
Taking a Cauchy sequence
$(\Phi)_m \in C^n_c(\mathbb{R}^3, {\cal G})$ converging to an element $Z \in
H_n(\mathbb{R}^3, {\cal G})$ proves the embedding (\ref{embed}) in the limit $m
\rightarrow \infty$. $\Box$

\subsection{Proposition \ref{prior}}\label{priorpr}

The proof is elementary in principle but tedious in practice. It consists of two parts,
the first being straightforward calculus with smooth enough gauge potentials, while in
the second part the smoothness assumptions are relaxed by examining the convergence
properties of the curvature norms. We begin by assuming that $A_k \in C^n(\mathbb{R}^3,
{\cal G})$ and then consider a mapping $\Psi \in C^3_c(\mathbb{R}^3, {\cal G})$. Making
use of the fact that
$$ [\nabla_k,\nabla_l]_{op} \, \Psi = - [G_{kl},\Psi], $$
where $[\cdot,\cdot]_{op}$ stands for the operator commutator of two covariant
derivatives, we deduce the identity
\begin{eqnarray*}
\lefteqn{w^{\beta} |\nabla_k \nabla_l \Psi|^2 = } \quad \\*
&& \hspace{-1.2em} \partial_l \left[ w^{\beta} (\nabla_k \Psi, \nabla_k \nabla_l \Psi)
\right] - \partial_k \left[ w^{\beta} (\nabla_k \Psi, \nabla_l \nabla_l \Psi) \right] -
\partial_k \left[ w^{\beta} ([G_{kl},\Psi], \nabla_l \Psi) \right] \\
&& \hspace{-1.2em} - \partial_k \left[ (\partial_l w^{\beta}) (\nabla_k \Psi, \nabla_l
\Psi) \right] + (\partial_k \partial_l \, w^{\beta}) (\nabla_k \Psi, \nabla_l \Psi) +
(\partial_l w^{\beta}) (\nabla_k \nabla_k \Psi, \nabla_l \Psi) \\
&& \hspace{-1.2em} + (\partial_k w^{\beta}) (\nabla_l \nabla_l \Psi, \nabla_k \Psi) +
(\partial_k w^{\beta}) ([G_{kl},\Psi], \nabla_l \Psi) + w^{\beta} ([\nabla_k G_{kl},\Psi],
\nabla_l \Psi) \\*
&& \hspace{-1.2em} - 2 w^{\beta} (\nabla_k \Psi, [G_{kl},\nabla_l \Psi]) + w^{\beta}
(\nabla_k \nabla_k \Psi, \nabla_l \nabla_l \Psi).
\end{eqnarray*}
Integrating this equation with the help of the Gauss--Green theorem, the surface terms
vanish due to the compact support of $\Psi$. Let us denote this support by $K$. Using
now Lemma \ref{kommu} and the inequality
$$ |\partial^k w^{\beta}| \leq C w^{\beta-k} $$
together with H\"{o}lder's inequality we find the estimate
\begin{eqnarray*}
\lefteqn{||w^{\beta/2} \nabla^2 \Psi||_2^2 \leq } \quad \\*
&& C \Bigl( ||w^{\beta/2 - 1} \nabla \Psi||_2^2 + ||w^{\beta/2} \Delta(A) \Psi||_2 \,
||w^{\beta/2 - 1} \nabla \Psi||_2 \\
&& \hspace{1.3em} + ||w^{2(1-\sigma)} G||_{\infty,K} \, ||w^{\beta/2 - (2-\sigma)}
\Psi||_2 \,
||w^{\beta/2 - (1-\sigma)} \nabla \Psi||_2 \\
&& \hspace{1.3em} + ||w^{2(1-\sigma)} (\nabla \cdot G)||_{3,K} \, ||w^{\beta/2 -
(1-\sigma)} \Psi||_6 \,
||w^{\beta/2 - (1-\sigma)} \nabla \Psi||_2 \\*
&& \hspace{1.3em} + ||w^{2(1-\sigma)} G||_{\infty,K} \, ||w^{\beta/2 - (1-\sigma)}
\nabla \Psi||_2^2 + ||w^{\beta/2} \Delta(A) \Psi||_2^2 \Bigr).
\end{eqnarray*}
Let us apply this inequality by choosing
$$ \Psi = \nabla_{l_1} \cdots \nabla_{l_{n-2}} \Phi, \qquad \beta = (2n-3)(1-\sigma). $$
If $n=2$, we employ Proposition \ref{goin}, inequality (\ref{gns2}), the inequality
\begin{equation}\label{apu72}
w^{\beta/2 - 1} \leq w^{(n-5/2)(1-\sigma)}
\end{equation}
and finally Cauchy's inequality to get
\begin{eqnarray}\label{apu73}
||w^{\frac{1-\sigma}{2}} \nabla^2 \Phi||_2^2 &\leq& C \Bigl\{ ||w^{\frac{1-\sigma}{2}}
\Delta(A) \Phi||_2^2 + \Bigl( 1 + ||w^{2(1-\sigma)} G||_{\infty,K} \nonumber \\*
&& \hspace{1.6em} + ||w^{2(1-\sigma)} (\nabla \cdot G)||_{3,K} \Bigr) \,
||w^{-\frac{1-\sigma}{2}} \nabla \Phi||_2^2 \Bigr\}.
\end{eqnarray}
This establishes a pre-stage of inequality (\ref{apri}), where the curvature norms are
local i.e. they are taken over the support of $\Phi$. The constant $C$ depends on
$\Phi$ only through these curvature norms. For higher order derivatives with $n \geq 3$
we use inequalities (\ref{gns2}) and (\ref{apu72}) to bound the derivative norms up to
the $(n-1)$th order. The highest order term with the covariant Laplacian is handled by
writing
$$ \nabla_k \nabla_k \nabla_{l_1} \cdots \nabla_{l_{n-2}} \Phi = \nabla_{l_1} \cdots
\nabla_{l_{n-2}} \nabla_k \nabla_k \Phi + [\nabla_k \nabla_k, \nabla_{l_1} \cdots
\nabla_{l_{n-2}}]_{op} \, \Phi $$
and employing the decomposition
\begin{eqnarray*}
\lefteqn{[\nabla_k \nabla_k, \nabla_{l_1} \cdots \nabla_{l_{n-2}}]_{op} \, \Phi = -
\sum_{p=0}^{n-3} \sum_{s_1=0}^1 \cdots \sum_{s_p=0}^1} \\*
&& \Bigl( [ \nabla_{l_1}^{s_1} \cdots \nabla_{l_p}^{s_p} \nabla_k G_{kl_{p+1}},
\nabla_{l_1}^{1-s_1} \cdots \nabla_{l_p}^{1-s_p} \nabla_{l_{p+2}} \cdots
\nabla_{l_{n-2}} \Phi] \\*
&& \hspace{0.7em} + 2 \, [ \nabla_{l_1}^{s_1} \cdots \nabla_{l_p}^{s_p} G_{kl_{p+1}},
\nabla_{l_1}^{1-s_1} \cdots \nabla_{l_p}^{1-s_p} \nabla_k \nabla_{l_{p+2}} \cdots
\nabla_{l_{n-2}} \Phi] \Bigr),
\end{eqnarray*}
which can be proved by induction. If $p=0$ or $p=n-3$ this formula requires some
interpretation, suppressing those sums and derivatives which become ill-defined. The
important thing is to notice that the terms of this decomposition, when summed over
$k$, are bounded in the Lie algebra norm by the norms
$$ |\nabla^p (\nabla \cdot G)| \, |\nabla^{n-3-p} \Phi|, \quad |\nabla^p G| \,
|\nabla^{n-2-p}\Phi|, \quad 0 \leq p \leq n-3. $$ Therefore, upon using H\"{o}lder's
and Cauchy's inequalities together with the estimate (\ref{gns2}), we deduce the bound
\begin{eqnarray}\label{apu74}
\lefteqn{||w^{(n-3/2)(1-\sigma)} \nabla^n \Phi||_2^2 \leq} \nonumber \\*
&& \hspace{1em} C \Bigl\{ ||w^{(n-3/2)(1-\sigma)} \nabla^{n-2} \Delta(A) \Phi||_2^2
\nonumber \\
&& \hspace{2.5em} + \sum_{p=0}^{n-3} \Bigl( ||w^{(p+2)(1-\sigma)} \nabla^p (\nabla
\cdot G)||_{3,K}^2 \, ||\Phi||_{n-2-p,2}^2 \\*
&& \hspace{5.6em} + ||w^{(p+2)(1-\sigma)} \nabla^p G||_{\infty,K}^2 \,
||\Phi||_{n-2-p,2}^2 \Bigr) \nonumber \\*
&& \hspace{2.5em} + \left( 1 + ||w^{2(1-\sigma)} G||_{\infty,K} + ||w^{2(1-\sigma)}
(\nabla \cdot G)||_{3,K} \right) \, ||\Phi||_{n-1,2}^2 \Bigr\}, \nonumber
\end{eqnarray}
which in turn leads to an inequality like (\ref{apri}) with local curvature norms.

We are now done with the first part of the proof. The requirement $A_k \in
C^n(\mathbb{R}^3, {\cal G})$ was dictated by the Gauss--Green theorem, but we can relax
this condition by a few orders of smoothness. In fact, I am going to prove that the
assumptions (\ref{asmo}) are sufficient for Proposition \ref{prior} to hold. The
technique is to apply the estimates (\ref{apu73}) and (\ref{apu74}) to mollified gauge
potentials. If the norms appearing in these inequalities converge to the original norms
in the limit when the mollification is removed, we have succeeded in proving the
inequalities with relaxed smoothness assumptions. Mollification is defined by the usual
formula
$$ A^{(\delta)}_k(x) = (\eta_{\delta}*A_k)(x) = \int \eta_{\delta}(x-y) A_k(y) \, d^3 y $$
with
$$ \eta_{\delta}(x) =  \frac{1}{\delta^3} \eta \left( \frac{x}{\delta} \right), \quad
\eta(x) = \left\{ \begin{array}{ll} C \exp \left( -\frac{1}{1 - |x|^2} \right), & |x| <
1 \\ 0, & |x| \geq 1, \end{array} \right. $$
where the constant $C$ is fixed so that
\begin{equation}\label{etano}
\int \eta(x) \, d^3 x = 1.
\end{equation}
It is also assumed that all derivatives of the gauge potential $A_k$ up to the highest
order implied by the curvature norms are at least locally integrable. This assumption
ensures that the order of differentiation and mollification can be reversed. The
covariant derivative related to the mollified potential will be denoted by
$$ \nabla_k^{(\delta)} = \partial_k + [A^{(\delta)}_k, \, \cdot \,], $$
and the corresponding curvature form is
$$ \widehat{G}_{kl}^{(\delta)} = \partial_l A^{(\delta)}_k - \partial_k A^{(\delta)}_l -
[A^{(\delta)}_k, A^{(\delta)}_l]. $$
I also define a mollified norm $|\nabla^p G|^{(\delta)}$
by
$$ |\nabla^p G|^{(\delta)} = \Bigl( \sum_{k_1} \cdots \sum_{k_p} \sum_{l} \sum_{m} \Bigl|
\int \eta_{\delta}(x-y) \nabla_{k_1} \cdots \nabla_{k_p} G_{lm}(y) \, d^3 y \Bigr|^2
\Bigr)^{1/2}. $$
It is well-known that in the limit $\delta \rightarrow 0$
\begin{equation}\label{apu81}
||w^{(p+2)(1-\sigma)} |\nabla^p G|^{(\delta)}||_{\infty,K} \longrightarrow
||w^{(p+2)(1-\sigma)} |\nabla^p G| \, ||_{\infty,K},
\end{equation}
when $\nabla^p G_{lm} \in C(\mathbb{R}^3, {\cal G})$. However, we eventually need the
result
\begin{equation}\label{apu82}
||w^{(p+2)(1-\sigma)} |(\nabla^{(\delta)})^p \widehat{G}^{(\delta)}| \, ||_{\infty,K}
\longrightarrow ||w^{(p+2)(1-\sigma)} |\nabla^p G| \, ||_{\infty,K},
\end{equation}
and this can be proved using the limit (\ref{apu81}) combined with the estimate
\begin{eqnarray}\label{apu83}
\lefteqn{ \hspace{-2em} \Bigl| \Bigl| \, w^{(p+2)(1-\sigma)} \left(
|\nabla^p G|^{(\delta)} - |(\nabla^{(\delta)})^p \widehat{G}^{(\delta)}| \right) \Bigr|
\Bigr|_{\, \infty,K} \leq \sum_{k_1} \cdots \sum_{k_p} \sum_{l} \sum_{m}} \\*
&& \hspace{-1em} \Bigl| \Bigl| \, w^{(p+2)(1-\sigma)} \bigl| \,
\eta_{\delta}*(\nabla_{k_1} \cdots \nabla_{k_p} G_{lm}) - \nabla_{k_1}^{(\delta)} \cdots
\nabla_{k_p}^{(\delta)} \widehat{G}_{lm}^{(\delta)} \bigr| \, \Bigr| \Bigr|_{\, \infty,K}.
\nonumber
\end{eqnarray}
Using the decomposition
\begin{eqnarray}\label{apu84}
&& \nabla_{k_1} \cdots \nabla_{k_p} G_{lm} = \sum_{j=0}^p \sum_{i_1=1}^{i_2-1} \cdots
\sum_{i_j=j}^{p} \partial_{k_1} \cdots \partial_{k_{i_1 -1}} (\textrm{ad}
A_{k_{i_1}}) \partial_{k_{i_1 +1}} \nonumber \\*
&& \hspace{11.0em} \cdots \partial_{k_{i_j -1}} (\textrm{ad} A_{k_{i_j}})
\partial_{k_{i_j +1}} \cdots \partial_{k_p} G_{lm}
\end{eqnarray}
with the notation $\textrm{ad} A_k = [A_k, \, \cdot \,]$ and recalling the definition
of the curvature form we see that each term of this decomposition takes the form
\begin{subequations}\label{apu85}
\begin{equation}\label{apu85a}
 \textrm{ad} (\partial^{\alpha_{i_1}} A_{k_{i_1}}) \cdots \textrm{ad}
(\partial^{\alpha_{i_j}} A_{k_{i_j}}) \partial^{\alpha_{i_{j+1}}} A_{k_{p+2}},
\end{equation}
where the multi-indices $\alpha_{i_j}$ satisfy
\begin{equation}\label{apu85b}
\sum_{s=1}^{j+1} |\alpha_{i_s}| = p+1-j, \qquad 0 \leq j \leq p+1.
\end{equation}
\end{subequations}
Since there is a finite number of terms in formulas (\ref{apu83}) and (\ref{apu84})
altogether, it suffices to consider each term separately. In inequality (\ref{apu83})
the contribution from each term of the form (\ref{apu85a}) is
\begin{subequations}\label{apu86}
\begin{eqnarray}
&& \Bigl| \Bigl| \, w^{(p+2)(1-\sigma)} \bigl| \, \eta_{\delta}* \left( \textrm{ad}
(\partial^{\alpha_{i_1}} A_{k_{i_1}}) \cdots \textrm{ad} (\partial^{\alpha_{i_j}}
A_{k_{i_j}})
\partial^{\alpha_{i_{j+1}}} A_{k_{p+2}} \right) \nonumber \\*
&& \hspace{6em} - \textrm{ad} (\partial^{\alpha_{i_1}} A_{k_{i_1}}^{(\delta)}) \cdots
\textrm{ad} (\partial^{\alpha_{i_j}} A_{k_{i_j}}^{(\delta)})
\partial^{\alpha_{i_{j+1}}} A_{k_{p+2}}^{(\delta)} \bigr| \, \Bigr| \Bigr|_{\, \infty,K}
\nonumber \\
&& \leq \textrm{ess}\sup_{\hspace{-1.3em} x \in K} \, C \, \Bigl| \int \! d^3 y_1
\cdots d^3 y_{j+1} \, \eta_{\delta}(x-y_1)
\cdots \eta_{\delta}(x-y_{j+1}) \label{apu86a} \\*
&& \hspace{6em} \Bigl\{ \textrm{ad} (\partial^{\alpha_{i_1}} A_{k_{i_1}}(y_{j+1}))
\cdots \textrm{ad} (\partial^{\alpha_{i_j}} A_{k_{i_j}}(y_{j+1}))
\partial^{\alpha_{i_{j+1}}} A_{k_{p+2}}(y_{j+1}) \nonumber \\*
&& \hspace{6.8em} - \textrm{ad} (\partial^{\alpha_{i_1}} A_{k_{i_1}}(y_{1})) \cdots
\textrm{ad} (\partial^{\alpha_{i_j}} A_{k_{i_j}}(y_j))
\partial^{\alpha_{i_{j+1}}} A_{k_{p+2}}(y_{j+1}) \Bigr\} \Bigr| \nonumber \\
&& \leq C \sum_{s=1}^j \textrm{ess}\sup_{\hspace{-1.3em} x \in K} \, \Bigl| \int \! d^3
y_1 \cdots d^3 y_{j+1} \, \eta_{\delta}(x-y_1)
\cdots \eta_{\delta}(x-y_{j+1}) \label{apu86b} \\*
&& \hspace{7.3em} \textrm{ad} (\partial^{\alpha_{i_1}} A_{k_{i_1}}(y_{j+1})) \cdots
\textrm{ad} \left( \partial^{\alpha_{i_s}} A_{k_{i_s}}(y_{j+1}) -
\partial^{\alpha_{i_s}} A_{k_{i_s}}(y_s) \right) \nonumber \\*
&& \hspace{7.3em}  \textrm{ad} (\partial^{\alpha_{i_{s+1}}} A_{k_{i_{s+1}}}(y_{s+1}))
\cdots \partial^{\alpha_{i_{j+1}}} A_{k_{p+2}}(y_{j+1}) \Bigr| \, , \qquad j \geq 1,
\nonumber
\end{eqnarray}
\end{subequations}
and in the case $j=0$ the expression (\ref{apu86a}) vanishes. The constant $C$
corresponds to the upper bound of the weight factor on $K$, and in deriving the last
inequality the formula
$$ \textrm{ad} B_1 \cdots \textrm{ad} B_m - \textrm{ad} C_1 \cdots \textrm{ad} C_m
= \sum_{s=1}^m \textrm{ad} B_1 \cdots \textrm{ad} (B_s - C_s) \textrm{ad} C_{s+1}
\cdots \textrm{ad} C_m $$ was used. Applying now H\"{o}lder's inequality and Lemma
\ref{kommu} to the expression (\ref{apu86b}) with the normalisation (\ref{etano}) in
mind, we see that each term in the sum is bounded by
\begin{eqnarray*}
&& \textrm{ess} \hspace{-0.7em} \sup_{\substack{\hspace{-1.7em} y_s, y_{j+1} \in
K_{\delta} \\ \hspace{-1.8em} |y_s - y_{j+1}| \leq \, 2\delta}} \left| \,
\partial^{\alpha_{i_s}} A_{k_{i_s}}(y_{j+1}) - \partial^{\alpha_{i_s}} A_{k_{i_s}}(y_s)
\right| \, \prod_{\substack{r=1 \\ r \neq s}}^{j+1} \textrm{ess}\sup_{\hspace{-1.3em} y
\in K_{\delta}} \left| \, \partial^{\alpha_{i_r}} A_{k_{i_r}}(y) \right|, \\
&& K_{\delta} = \left\{ x \in \mathbb{R}^3 \, | \, \text{dist}(x,K) \leq \delta
\right\}.
\end{eqnarray*}
Since now
$$ |\alpha_{i_s}| \leq p \leq p_{max} = \left\{ \begin{array}{ll} 0, & \quad n = 2 \\
n-3, & \quad n \geq 3, \end{array} \right. $$
the bound above tends to zero in the
limit $\delta \rightarrow 0$ for gauge potentials of class $C^{p_{max}}(\mathbb{R}^3,
{\cal G})$. As a result,
$$ \Bigl| \Bigl| \, w^{(p+2)(1-\sigma)} \left( |\nabla^p G|^{(\delta)} -
|(\nabla^{(\delta)})^p \widehat{G}^{(\delta)}| \right) \Bigr| \Bigr|_{\, \infty,K}
\longrightarrow \, 0, $$
and together with the limit (\ref{apu81}) this establishes the
desired convergence property (\ref{apu82}). The $L^3$ curvature norm is treated
similarly. We define a mollified norm
\begin{eqnarray*}
\lefteqn{|\nabla^p (\nabla \cdot G)|^{(\delta)} =} \\*
&& \Bigl( \sum_{k_1} \cdots \sum_{k_p} \sum_{m} \Bigl| \int \eta_{\delta}(x-y)
\nabla_{k_1} \cdots \nabla_{k_p} \sum_{l} \nabla_l G_{lm}(y) \, d^3 y \Bigr|^2
\Bigr)^{1/2},
\end{eqnarray*}
which satisfies
\begin{equation}\label{apu87}
||w^{(p+2)(1-\sigma)} |\nabla^p (\nabla \cdot G)|^{(\delta)}||_{3,K} \xrightarrow[\delta
\rightarrow 0]{} ||w^{(p+2)(1-\sigma)} |\nabla^p (\nabla \cdot G)| \, ||_{3,K}
\end{equation}
when $|\nabla^p (\nabla \cdot G)| \in L^3_{loc}$. An estimate for the norm
\begin{equation}\label{apu88}
\Bigl| \Bigl| \, w^{(p+2)(1-\sigma)} \left( |\nabla^p (\nabla \cdot G)|^{(\delta)} -
|(\nabla^{(\delta)})^p (\nabla^{(\delta)} \cdot \widehat{G}^{(\delta)})| \right) \Bigr|
\Bigr|_{\, 3,K}
\end{equation}
is obtained by decomposing the expression
$$ \nabla_{k_1} \cdots \nabla_{k_p} \sum_{l} \nabla_l G_{lm} $$
into terms of the form (\ref{apu85}) with $p$ replaced by $p+1$ as there is one
additional derivative now. As before, only terms with $j \geq 1$ contribute to the norm
(\ref{apu88}), these contributions being bounded by the integrals
\begin{eqnarray}\label{apu89}
&& \int_K \! d^3 x \, \Bigl| \, \int \! d^3 y_1 \cdots d^3 y_{j+1} \,
\eta_{\delta}(y_1) \cdots \eta_{\delta}(y_{j+1}) \nonumber \\*
&& \hspace{1.8em} \textrm{ad} (\partial^{\alpha_{i_1}} A_{k_{i_1}}(x-y_{j+1})) \cdots
\textrm{ad} \left( \partial^{\alpha_{i_s}} A_{k_{i_s}}(x-y_{j+1}) -
\partial^{\alpha_{i_s}} A_{k_{i_s}}(x-y_s) \right) \nonumber \\*
&& \hspace{1.8em}  \textrm{ad} (\partial^{\alpha_{i_{s+1}}} A_{k_{i_{s+1}}}(x-y_{s+1}))
\cdots \partial^{\alpha_{i_{j+1}}} A_{k_{p+3}}(x-y_{j+1}) \Bigr|^{\, 3} \nonumber \\
&& \leq C \, \textrm{ess}\sup_{\hspace{-1.3em} |y_1| \leq \, \delta} \cdots
\textrm{ess}\sup_{\hspace{-1.6em} |y_{j+1}| \leq \, \delta} \\*
&& \hspace{0.6em} \int_K \! d^3 x \, \Bigl( \left| \, \partial^{\alpha_{i_1}}
A_{k_{i_1}}(x-y_{j+1}) \right| \cdots \left| \, \partial^{\alpha_{i_s}}
A_{k_{i_s}}(x-y_{j+1}) - \partial^{\alpha_{i_s}} A_{k_{i_s}}(x-y_s) \right| \nonumber \\*
&& \hspace{4.2em} \bigl| \, \partial^{\alpha_{i_{s+1}}} A_{k_{i_{s+1}}}(x-y_{s+1})
\bigr| \cdots \left| \, \partial^{\alpha_{i_{j+1}}} A_{k_{p+3}}(x-y_{j+1}) \right|
\Bigr)^3 . \nonumber
\end{eqnarray}
The last inequality was derived by employing Lemma \ref{kommu} and H\"{o}lder's
inequality repeatedly. Now if $j \geq 2$, we have $|\alpha_{i_s}| \leq p_{max}$ and the
integrand of the estimate (\ref{apu89}) is continuous, meaning that the whole
expression vanishes in the limit $\delta \rightarrow 0$. In the case $j=1$ there are
two multi-indices satisfying
$$|\alpha_{i_1}| + |\alpha_{i_2}| \leq p_{max} + 1.$$
If one of them takes the maximal value, the other vanishes and accordingly, we can make
use of the fact that $\partial^{p_{max} + 1} A_k \in L^3_{loc}(\mathbb{R}^3, {\cal G})$
and employ H\"{o}lder's inequality to get such a bound for the integral (\ref{apu89})
that it vanishes when $\delta$ is sent to zero. As a result, the norm (\ref{apu88})
also has a vanishing limit, and combined with the limit (\ref{apu87}) we get
\begin{equation}\label{apu810}
||w^{(p+2)(1-\sigma)} |(\nabla^{(\delta)})^p ( \nabla^{(\delta)} \cdot
\widehat{G}^{(\delta)})| \, ||_{3,K} \longrightarrow ||w^{(p+2)(1-\sigma)} |\nabla^p
(\nabla \cdot G)| \, ||_{3,K}.
\end{equation}
It still remains to consider the convergence of the Sobolev norms. Let us denote by
$||\Phi||_{n,2}^{(\delta)}$ the norm evaluated with the mollified gauge potential. From
the estimate
\begin{eqnarray}\label{apu91}
\lefteqn{ \left| \, ||\Phi||_{n,2} - ||\Phi||_{n,2}^{(\delta)} \, \right| \leq} \\*
&& \sum_{p=1}^n \sum_{k_1} \cdots \sum_{k_p} \, \Bigl| \Bigl| w^{(p-3/2)(1-\sigma)}
\bigl| \nabla_{k_1} \cdots \nabla_{k_p} \Phi - \nabla_{k_1}^{(\delta)} \cdots
\nabla_{k_p}^{(\delta)} \Phi \bigr| \, \Bigr| \Bigr|_2 \nonumber
\end{eqnarray}
and the decomposition of the derivatives
$$ \nabla_{k_1} \cdots \nabla_{k_p} \Phi $$
into terms of the form
\begin{eqnarray*}
&& \textrm{ad} (\partial^{\alpha_{i_1}} A_{k_{i_1}}) \cdots \textrm{ad}
(\partial^{\alpha_{i_j}} A_{k_{i_j}}) \partial^{\alpha_{i_{j+1}}} \Phi, \\
&& \sum_{s=1}^{j+1} |\alpha_{i_s}| = p-j, \qquad 0 \leq j \leq p
\end{eqnarray*}
it is evident that each term in inequality (\ref{apu91}) is bounded by integrals of the
form
\begin{eqnarray}\label{apu93}
&& \int_K \! d^3 x \, \Bigl| \, \int \! d^3 y_1 \cdots d^3 y_j \,
\eta_{\delta}(y_1) \cdots \eta_{\delta}(y_j) \nonumber \\*
&& \hspace{3.4em} \textrm{ad} (\partial^{\alpha_{i_1}} A_{k_{i_1}}(x)) \cdots
\textrm{ad} \left( \partial^{\alpha_{i_s}} A_{k_{i_s}}(x) -
\partial^{\alpha_{i_s}} A_{k_{i_s}}(x-y_s) \right) \nonumber \\*
&& \hspace{3.4em}  \textrm{ad} (\partial^{\alpha_{i_{s+1}}} A_{k_{i_{s+1}}}(x-y_{s+1}))
\cdots \partial^{\alpha_{i_{j+1}}} \Phi(x) \Bigr|^{\, 2} \nonumber \\
&& \leq C \, \textrm{ess}\sup_{\hspace{-1.3em} |y_1| \leq \, \delta} \cdots
\textrm{ess}\sup_{\hspace{-1.3em} |y_j| \leq \, \delta} \\*
&& \hspace{2.4em} \int_K \! d^3 x \, \Bigl( \left| \, \partial^{\alpha_{i_1}}
A_{k_{i_1}}(x) \right| \cdots \left| \, \partial^{\alpha_{i_s}}
A_{k_{i_s}}(x) - \partial^{\alpha_{i_s}} A_{k_{i_s}}(x-y_s) \right| \nonumber \\*
&& \hspace{6em} \bigl| \, \partial^{\alpha_{i_{s+1}}} A_{k_{i_{s+1}}}(x-y_{s+1}) \bigr|
\cdots \left| \, \partial^{\alpha_{i_{j+1}}} \Phi(x) \right| \Bigr)^2 \nonumber
\end{eqnarray}
with $j \geq 1$. Inspection of all possible values of the multi-indices
$|\alpha_{i_s}|$ indicates that the previous smoothness assumptions, complemented by
$\partial^{n-1} A_k \in L^2_{loc}(\mathbb{R}^3, {\cal G})$, are sufficient to make the
bound (\ref{apu93}) vanish in the limit $\delta \rightarrow 0$. What this means is that
$$ ||\Phi||_{n,2}^{(\delta)} \longrightarrow ||\Phi||_{n,2}, $$
and together with the limits (\ref{apu82}) and (\ref{apu810}) this completes the proof
of inequalities (\ref{apu73}) and (\ref{apu74}) under the relaxed assumptions
(\ref{asmo}). Replacing the local curvature norms by global ones we finally find
ourselves at the end of this long proof. $\Box$

\subsection{Lemma \ref{jono}}\label{jonopr}

Intuitively this lemma is rather obvious, but strictly speaking we do not yet know that
a smooth enough sequence converging in $H_{n-1}(\mathbb{R}^3, {\cal G})$ also converges
in the norm involving the covariant Laplacian. The proof follows the usual steps,
beginning with approximations with truncated mappings and ending with density results
for mollified mappings. A sequence of truncated mappings is defined by
$$ Z_m(x) = \tau_m(x) Z(x) $$
with
$$ \tau_m(x) = \tau \left( \frac{x}{m} \right), \quad \tau(x) = \left\{
\begin{array}{ll} 1, & \quad 0 \leq |x| \leq 1 \\
0, & \quad |x| \geq 2. \end{array} \right. $$
When $1 \leq |x| \leq 2$, it is assumed
that $\tau$ is infinitely smooth and $0 \leq \tau(x) \leq 1$. In order to approximate
the Sobolev norm we notice that
\begin{equation}\label{apu101}
||Z_m - Z||_{n-1,2} \leq \sum_{p=1}^{n-1} \sum_{k_1} \cdots \sum_{k_p} ||
w^{(p-3/2)(1-\sigma)} \nabla_{k_1} \cdots \nabla_{k_p} (Z_m - Z) ||_2
\end{equation}
and then estimate each term in the sum separately. The covariant derivatives are
decomposed as
\begin{equation}\label{apu102}
\nabla_{k_1} \cdots \nabla_{k_p} (Z_m - Z) = (\tau_m - 1) \nabla_{k_1} \cdots
\nabla_{k_p} Z + \sideset{}{'} \sum_{|\alpha| \leq p-1} (\partial^{\alpha_c} \tau_m) \,
\nabla^{\alpha} Z,
\end{equation}
where the prime indicates that the indices of $\alpha$ are assumed to be in ascending
order,
$$ \alpha = (k_{i_1},\dots, k_{i_q}), \quad 1 \leq i_1 < \cdots < i_q \leq p $$
and where $\alpha_c$ stands for the complement of $\alpha$, i.e.
$$ \alpha_c = (k_1,\dots,k_p) \backslash \alpha. $$
The norm of the first term in the decomposition (\ref{apu102}) can be made small by
using the result that
\begin{equation}\label{apu103}
\int_{|x| \geq \, m} w^{(2p-3)(1-\sigma)} |\nabla^p Z|^2 \, d^3 x \xrightarrow[m
\rightarrow \infty]{} 0,
\end{equation}
when $Z \in H_{n-1}(\mathbb{R}^3, {\cal G})$. For the remaining terms with $|\alpha|
\geq 1$ we deduce the bound
\begin{eqnarray}\label{apu104}
&& \int w^{(2p-3)(1-\sigma)} |\partial^{\alpha_c} \tau_m|^2 \, |\nabla^{\alpha} Z|^2 \,
d^3 x \leq \sup_{1 \leq \frac{|x|}{m} \leq 2} w^{2(p-|\alpha|)(1-\sigma)}
|\partial^{\alpha_c} \tau_m|^2 \nonumber \\*
&& \hspace{1em} \times \int_{1 \leq \frac{|x|}{m} \leq 2} w^{(2|\alpha|-3)(1-\sigma)}
|\nabla^{\alpha} Z|^2 \, d^3 x.
\end{eqnarray}
The factor on the right hand side is bounded, because
\begin{eqnarray*}
&& \sup_{1 \leq \frac{|x|}{m} \leq \, 2} \hspace{-0.6em} w(x)^{2(p-|\alpha|)(1-\sigma)}
|\partial^{\alpha_c} \tau_m(x)|^2 \leq \\*
&& \hspace{4em} m^{-2(p-|\alpha|)\sigma} \sup_{1 \leq \xi \leq \, 2} \hspace{-0.2em}
w(\xi)^{2(p-|\alpha|)(1-\sigma)} |\partial^{\alpha_c} \tau(\xi)|^2, \qquad \xi =
\frac{x}{m},
\end{eqnarray*}
and the integral in formula (\ref{apu104}) tends to zero by virtue of the limit
(\ref{apu103}). Similarly, the term with $|\alpha| = 0$, $\alpha_c = (k_1,\dots,k_p)$
is seen to converge to zero by the estimates
\begin{eqnarray*}
&& \int w^{(2p-3)(1-\sigma)} |\partial^{\alpha_c} \tau_m|^2 \, |Z|^2 \, d^3 x \leq
\sup_{1
\leq \frac{|x|}{m} \leq 2} w^{2p - 2(p-1)\sigma} |\partial^{\alpha_c} \tau_m|^2 \\*
&& \hspace{1em} \times \int_{1 \leq \frac{|x|}{m} \leq 2} w^{-(3-\sigma)} |Z|^2 \, d^3
x
\end{eqnarray*}
and
\begin{eqnarray*}
&& \sup_{1 \leq \frac{|x|}{m} \leq \, 2} \hspace{-0.6em} w(x)^{2p - 2(p-1)\sigma}
|\partial^{\alpha_c} \tau_m(x)|^2 \leq \\*
&& \hspace{4em} m^{-2(p-1)\sigma} \sup_{1 \leq \xi \leq \, 2} \hspace{-0.2em}
w(\xi)^{2p - 2(p-1)\sigma} |\partial^{\alpha_c} \tau(\xi)|^2.
\end{eqnarray*}
The norm with the covariant Laplacian is handled similarly, employing the decomposition
\begin{eqnarray*}
\nabla_{k_1} \cdots \nabla_{k_{n-2}} \Delta(A) (Z_m - Z) &=& (\tau_m - 1) \nabla_{k_1}
\cdots \nabla_{k_{n-2}} \Delta(A) Z \\*
&& + \sum_l \sideset{}{'} \sum_{|\alpha| \leq n-1} (\partial^{\widehat{\alpha}_c}
\tau_m) \, \nabla^{\alpha} Z
\end{eqnarray*}
with
$$ \widehat{\alpha}_c = (k_1,\dots,k_{n-2},l,l) \backslash \alpha. $$
The first term yields a norm tending to zero due to the assumption (\ref{oljo}), and
the other terms are made small by the limit (\ref{apu103}). Since there is a finite
number of terms in both decompositions, we see that it is possible to choose a
truncated mapping $Z_m$ arbitrarily close to $Z$ in both norms.

The mapping $Z_m$ above vanishes (modulo sets of measure zero) outside some compact set
$K$. Therefore we can define a mapping of class $C^{\infty}_c(\mathbb{R}^3, {\cal G})$
by
$$ \Phi^{(\delta)}(x) = \int \eta_{\delta}(y) Z_m(x-y) \, d^3 y. $$
Let us now estimate the distance of these mappings from $Z_m$ in the two norms in
question. Making use of formula (\ref{apu101}) with $Z$ replaced by $\Phi^{(\delta)}$ we
are led to considering weighted $L^2$ norms of the derivatives $|\nabla^p
(\Phi^{(\delta)} - Z_m)|$. For this purpose we write
\begin{eqnarray*}
&& \nabla_k^{(x)} Z_m(x-y) = \left( \nabla_k^{(x-y)} + \textrm{ad} {\cal A}_k(x,y)
\right) Z_m(x-y), \\*
&& {\cal A}_k(x,y) = A_k(x) - A_k(x-y)
\end{eqnarray*}
and derive a decomposition similar to equation (\ref{apu84})
\begin{eqnarray}\label{apu105}
&& \nabla_{k_1}^{(x)} \cdots \nabla_{k_p}^{(x)} Z_m(x-y) = \sum_{j=0}^p
\sum_{i_1=1}^{i_2-1} \cdots \sum_{i_j=j}^{p} \nabla_{k_1}^{(x-y)} \cdots \nabla_{k_{i_1
-1}}^{(x-y)} (\textrm{ad} {\cal A}_{k_{i_1}}) \nabla_{k_{i_1 +1}}^{(x-y)} \nonumber \\*
&& \hspace{12.0em}  \cdots \nabla_{k_{i_j -1}}^{(x-y)} (\textrm{ad} {\cal A}_{k_{i_j}})
\nabla_{k_{i_j +1}}^{(x-y)} \cdots \nabla_{k_p}^{(x-y)} Z_m(x-y) \nonumber \\*
&& \hspace{2em} = \nabla_{k_1}^{(x-y)} \cdots \nabla_{k_p}^{(x-y)} Z_m(x-y) + \,
\cdots,
\end{eqnarray}
where the remaining terms take the form
\begin{subequations}\label{apu106}
\begin{equation}\label{apu106a}
 \textrm{ad} (\nabla^{\alpha_{i_1}}_{(x-y)} {\cal A}_{k_{i_1}}) \cdots \textrm{ad}
(\nabla^{\alpha_{i_j}}_{(x-y)} {\cal A}_{k_{i_j}}) \nabla^{\alpha_{i_{j+1}}}_{(x-y)}
Z_m(x-y)
\end{equation}
with
\begin{equation}\label{apu106b}
\sum_{s=1}^{j+1} |\alpha_{i_s}| = p-j \, , \qquad 1 \leq j \leq p \, .
\end{equation}
\end{subequations}
The first term of the decomposition (\ref{apu105}) is combined with the $p$th covariant
derivative of $Z_m(x)$ to get a bound
\begin{eqnarray*}
&& \int_{K_{\delta}} \! d^3 x \, w(x)^{(2p-3)(1-\sigma)} \Bigl| \, \int \! d^3 y \,
\eta_{\delta}(y) \Bigl( \nabla_{k_1}^{(x-y)} \cdots \nabla_{k_p}^{(x-y)} Z_m(x-y) \\*
&& \hspace{15.7em} - \nabla_{k_1}^{(x)} \cdots \nabla_{k_p}^{(x)} Z_m(x) \Bigr)
\Bigr|^{\, 2} \\
&& \leq C \, \textrm{ess} \sup_{\hspace{-1.3em} |y| \leq \, \delta} \int_{K_{\delta}}
\! d^3 x \, \left| \, \nabla_{k_1}^{(x-y)} \cdots \nabla_{k_p}^{(x-y)} Z_m(x-y) -
\nabla_{k_1}^{(x)} \cdots \nabla_{k_p}^{(x)} Z_m(x) \right|^{\, 2},
\end{eqnarray*}
which tends to zero in the limit $\delta \rightarrow 0$, because now $\nabla^p Z_m \in
L^2_{loc}(\mathbb{R}^3, {\cal G})$. For terms of the form (\ref{apu106}) we get bounds
of the form
$$ \textrm{ess} \sup_{\hspace{-1.3em} |y| \leq \, \delta} \int_{K_{\delta}} \! d^3 x \,
\left( |\nabla^{\alpha_{i_1}}_{(x-y)} {\cal A}_{k_{i_1}}| \cdots
|\nabla^{\alpha_{i_j}}_{(x-y)} {\cal A}_{k_{i_j}}| \,
|\nabla^{\alpha_{i_{j+1}}}_{(x-y)} Z_m(x-y)| \right)^2. $$
Making use of H\"{o}lder's
inequality and the fact that $Z_m$ is continuous for $n \geq 3$ we can check that the
conditions (\ref{asmo}) are sufficient to make this integral converge to zero as $\delta
 \rightarrow 0$. The Laplacian norm is treated identically, writing
$$ \nabla_{k_1}^{(x)} \cdots \nabla_{k_{n-2}}^{(x)} \Delta^{(x)}(A) Z_m(x-y) =
\nabla_{k_1}^{(x-y)} \cdots \nabla_{k_{n-2}}^{(x-y)} \Delta^{(x-y)}(A) Z_m(x-y) + \,
\cdots, $$
where the remaining terms take the form (\ref{apu106}) with $p$ replaced by
$n$. For the first term we use the property that $\nabla^{n-2} \Delta(A) Z_m \in
L^2_{loc}(\mathbb{R}^3, {\cal G}) $ and for the others the properties
$$ Z_m \in L^6_{loc}(\mathbb{R}^3, {\cal G}) \cap C(\mathbb{R}^3, {\cal G}) \quad
\textrm{and} \quad \nabla^1 Z_m \in L^6_{loc}(\mathbb{R}^3, {\cal G}), \quad n \geq 3.
$$
There being only a finite number of terms, we can thus always find a mapping
$\Phi^{(\delta)}$ arbitrarily close to $Z_m$ in both norms. Accordingly, we can construct
a sequence $(\Phi_m)$ converging to $Z$ in these norms. $\Box$

\begin{acknowledgements}
I would like to thank professor C. Cronstr\"{o}m for fruitful discussions. In
particular, his results concerning the covariant Laplace equation inspired me to
consider weighted Sobolev spaces. This research was supported by the Academy of
Finland.
\end{acknowledgements}

\end{document}